\documentclass[twocolumn,preprintnumbers,showkeys,amsmath,amssymb]{revtex4}
\usepackage{graphicx}
\usepackage{amssymb}

\begin{document}
\title{Insecurity of a relativistic quantum commitment scheme}
\author{Guang Ping He}
\email{hegp@mail.sysu.edu.cn}
\affiliation{School of Physics, Sun Yat-sen University, Guangzhou 510275, China}

\begin{abstract}
We propose a cheating strategy to a relativistic quantum commitment scheme
[Sci Rep 2014;4:6774] which was claimed to be unconditionally secure. It is
shown that the sender Alice can cheat successfully with probability $100\%$,
thus disproving the security claim.
\end{abstract}


\maketitle



\section*{Introduction}

Quantum bit commitment (QBC) is an essential primitive of cryptography. It
is closely related with coin tossing and oblivious transfer, together they
form the base of multi-party secure computations. It is still widely
believed that non-relativistic unconditionally secure QBC is impossible \cite%
{qi24,qi23}, despite of the existence of counterexamples (e.g., Refs. \cite%
{HePRA,HeJPA,HeQIP,HeBook,HePRSA} and the references therein). Therefore,
some researches seek to improve the security with relativistic constraints
\cite{qi44,qi582,qbc24,qbc51,qbc66,qbc100}.

In these relativistic schemes, the participants are split into
\textquotedblleft agents\textquotedblright , so that it is no longer a
2-party cryptographic task. This makes it much easier to prevent the
participants from cheatings. But if the scheme is not properly designed,
cheatings could still be possible even in this multi-party scenario. Here we
will show that the scheme in Ref. \cite{qbc100} is such an example.

\section*{The scheme}

The original QBC is defined as a cryptographic task between two parties
Alice and Bob. It can be divided into two phases. In the commit phase, Alice
decides the value of the bit $b$ that she wants to commit, and sends Bob
some evidence, e.g., quantum states. Later, in the revealing phase, Alice
announces the value of $b$, and Bob checks it with the evidence. An
unconditionally secure QBC scheme needs to be both binding (i.e., Alice
cannot change the value of $b$ after the commit phase) and concealing (Bob
cannot know $b$ before the revealing phase).

The relativistic commitment scheme proposed in Ref. \cite{qbc100} is carried
on among three parties, the sender Alice, the receiver Bob and his agent C
(who is separated far away from Bob so that they cannot communicate
efficiently during the commit phase since no signal can travel faster than
the speed of light). The coding method of the scheme is as follows. Denote
the Bell states as (Eq. (1) of Ref. \cite{qbc100})%
\begin{equation}
\left\vert u_{i}u_{j}\right\rangle =\frac{\left\vert 0\right\rangle
\left\vert u_{j}\right\rangle +(-1)^{u_{i}}\left\vert 1\right\rangle
\left\vert 1\oplus u_{j}\right\rangle }{\sqrt{2}}.
\end{equation}%
To make a commitment, Alice secretly prepares $N$ Bell pairs $\left\vert
\beta _{ac}\right\rangle =\bigotimes\nolimits_{n=1}^{N}\left\vert
u_{a}u_{c}\right\rangle _{n}$\ and sends the second qubit of each pair to
agent C, while keeping the first qubit of each pair to herself. The values
of $u_{a}$ and $u_{c}$\ are chosen according to the value she wants to
commit. To commit to the bit $b=0$, Alice initiates with $\left\vert \beta
_{ac}\right\rangle =\bigotimes\nolimits_{n=1}^{N}\left\vert 00\right\rangle
_{n}$; to commit to $b=1$, Alice initiates with $\left\vert \beta
_{ac}\right\rangle =\bigotimes\nolimits_{n=1}^{N}\left\vert 01\right\rangle
_{n}$. Besides $b=0,1$, this scheme also allows Alice to commit to a qubit.
To commit to the qubit $(\left\vert 0\right\rangle +\left\vert
1\right\rangle )/\sqrt{2}$, Alice prepares $\left\vert \beta
_{ac}\right\rangle =\bigotimes\nolimits_{n=1}^{N}\left\vert 10\right\rangle
_{n}$; while for committing $(\left\vert 0\right\rangle -\left\vert
1\right\rangle )/\sqrt{2}$, Alice starts with $\left\vert \beta
_{ac}\right\rangle =\bigotimes\nolimits_{n=1}^{N}\left\vert 11\right\rangle
_{n}$.

Bob and agent C then perform some operations on their half of $\left\vert
\beta _{ac}\right\rangle $ and some other qubits at their side, which are
basically parts of the quantum teleportation process \cite{qi179}. But the
details of these operations are not important to us, because we shall show
that Alice can cheat anyway regardless what Bob and agent C do at their side.

In the revealing phase, Alice reveals her commitment by announcing the
values of $u_{a}u_{c}$. Also, to justify that she is honest, she sends to
Bob the other half of $\left\vert \beta _{ac}\right\rangle $ that she kept.

\section*{The cheating strategy}

Though the above scheme was claimed to be unconditionally secure in Ref.
\cite{qbc100}, here we show that a dishonest Alice can always alter her
commitment in the revealing phase by applying local operations on the qubits
she kept, while passing the security checks of Bob and agent C with
probability $100\%$.

Her cheating strategy is simple. From Eq. (1) we can see that the Pauli
matrix $\sigma _{z}$ acting on the first qubit alone can flip the value of
the bit $u_{i}$, while $(-1)^{u_{i}}\sigma _{x}$ acting on the first qubit
alone can flip the value of the bit $u_{j}$, i.e.,%
\begin{eqnarray}
(\sigma _{z}\otimes I)\left\vert u_{i}u_{j}\right\rangle &=&\left\vert \bar{u%
}_{i}u_{j}\right\rangle ,  \nonumber \\
((-1)^{u_{i}}\sigma _{x}\otimes I)\left\vert u_{i}u_{j}\right\rangle
&=&\left\vert u_{i}\bar{u}_{j}\right\rangle .
\end{eqnarray}%
Here $I$ is the identity operator on the second qubit. That is, Alice can
change $\left\vert u_{a}u_{c}\right\rangle _{n}$\ among the four Bell states
freely with a local unitary transformation\ of her own, without the help of
Bob and agent C.

Therefore, Alice can always starts the commitment phase with $\left\vert
\beta _{ac}\right\rangle =\bigotimes\nolimits_{n=1}^{N}\left\vert
00\right\rangle _{n}$. Later, in the revealing phase, if she wants to
convince Bob and agent C that she has committed to the bit $0$, she simply
follows the original scheme honestly. But if she wants to show that she has
committed to the bit $1$ (or the qubits $(\left\vert 0\right\rangle
+\left\vert 1\right\rangle )/\sqrt{2}$ or $(\left\vert 0\right\rangle
-\left\vert 1\right\rangle )/\sqrt{2}$), she applies $\sigma _{x}$\ ($\sigma
_{z}$ or $\sigma _{z}\sigma _{x}$) on each of the $N$ qubits she kept,
before sending them to B. Note that the local operations on the two qubits
of a Bell state, respectively, always commute with each other. Therefore, no
matter what operations Bob and agent C had performed on their share of the
qubits, Alice's above operation will make $\left\vert \beta
_{ac}\right\rangle $\ appear as if it was originally prepared as $%
\bigotimes\nolimits_{n=1}^{N}\left\vert 01\right\rangle _{n}$\ ($%
\bigotimes\nolimits_{n=1}^{N}\left\vert 10\right\rangle _{n}$ or $%
\bigotimes\nolimits_{n=1}^{N}\left\vert 11\right\rangle _{n}$), without
causing any conflict with the measurement results of Bob and agent C. Thus
her cheating can never be detected.

To summarize, the commitment scheme in Ref. \cite{qbc100} cannot meet the
binding condition against Alice's cheating, so that it is not
unconditionally secure.



\bigskip

\noindent \textbf{Funding:}

This work was supported in part by Guangdong
Basic and Applied Basic Research Foundation, China (Grant No. 2019A1515011048).





\begin{thebibliography}{99}
\bibitem{qi24} Mayers D. Unconditionally secure quantum bit commitment is
impossible. Phys Rev Lett 1997;78:3414.

\bibitem{qi23} Lo H-K, Chau HF. Is quantum bit commitment really possible?
Phys Rev Lett 1997;78:3410.

\bibitem{HePRA} He\ GP. Secure quantum bit commitment against empty
promises. Phys Rev A 2006;74:022332.

\bibitem{HeJPA} He\ GP. Quantum key distribution based on orthogonal states
allows secure quantum bit commitment. J Phys A 2011;44:445305.

\bibitem{HeQIP} He\ GP. Simplified quantum bit commitment using single
photon nonlocality. Quantum Inf Process 2014;13:2195.

\bibitem{HeBook} He GP. Chapter 4: Density matrices in quantum bit
commitment. In: N.V. Danielsen, editor. Understanding density matrices. New
York: Nova Science Publishers; 2019, p. 139--164.

\bibitem{HePRSA} He\ GP. Unconditionally secure quantum bit commitment based
on the uncertainty principle. Proc Roy Soc A 2019;475:20180543.

\bibitem{qi44} Kent A. Unconditionally secure bit commitment.\ Phys Rev Lett
1999;83:1447.

\bibitem{qi582} Kent A. Secure classical bit commitment using fixed capacity
communication channels.\ J Cryptol 2005;18:313.

\bibitem{qbc24} Kent A. Unconditionally secure bit commitment with flying
qudits.\ New J Phys 2011;13:113015.

\bibitem{qbc51} Kent A. Unconditionally secure bit commitment by
transmitting measurement outcomes.\ Phys Rev Lett 2012;109:130501.

\bibitem{qbc66} Kaniewski J, Tomamichel M, H\"{a}nggi E, Wehner S. Secure
bit commitment from relativistic constraints. IEEE Trans Inf Theory
2013;59:4687.

\bibitem{qbc100} Nadeem M. Unconditionally secure commitment in
position-based quantum cryptography. Sci Rep 2014;4:6774.

\bibitem{qi179} Bennett CH, Brassard G, Cr\'{e}peau C, Jozsa R, Peres A,
Wootters WK. Teleporting an unknown quantum state via dual classical and
Einstein-Podolsky-Rosen channels. Phys Rev Lett 1993;70:1895.
\end{thebibliography}
\end{document}